\documentclass[journal,twoside,print]{ieeecolor}
\usepackage{jsen}
\usepackage{cite}
\usepackage{amsmath,amssymb,amsfonts}
\usepackage{url}
\usepackage{booktabs} 
\usepackage{makecell} 
\usepackage{multirow}
\usepackage{algorithmic}
\usepackage{graphicx}
\usepackage{textcomp}
\usepackage{cite}
\usepackage{wrapfig}
\usepackage[a4paper, total={184mm,239mm}]{geometry}
\def\BibTeX{{\rm B\kern-.05em{\sc i\kern-.025em b}\kern-.08em
    T\kern-.1667em\lower.7ex\hbox{E}\kern-.125emX}}
\definecolor{subsectioncolor}{rgb}{0,0,0}
\definecolor{abstractbg}{rgb}{0.89804,0.94510,0.83137}
\makeatletter
\def\ps@titlepagestyle{\ps@headings}
\makeatother
\makeatletter

\renewcommand{\fnum@table}{%
  {\color{subsectioncolor}\sf Table~\thetable}%
}

\let\orig@makecaption\@makecaption

\long\def\@makecaption#1#2{%
  {\color{black}%
    \orig@makecaption{#1}{#2}%
  }%
 
}

\renewcommand{\fnum@figure}{%
  {\color{subsectioncolor}\sf Fig.~\thefigure}%
}

\makeatother

\makeatletter
\def\@seccntformat#1{\textcolor{subsectioncolor}{\csname the#1dis\endcsname}\hskip 0.5em\relax}
\makeatother

\begin{document}
\title{Transformer-based Neural Beamforming\\ for Real-Time Speech Enhancement\\ on Smart Low-Power Hearable Devices \\

}

\author{Luca Bompani\textsuperscript{1}, \IEEEmembership{Student Member, IEEE},
Marco Fariselli, Giovanni Oltrecolli\textsuperscript{1},
Francesco Conti\textsuperscript{1}, \IEEEmembership{Senior Member, IEEE}\\
\textsuperscript{1}Electrical, Electronic and Information Engineering (DEI), University of Bologna, Italy}

\IEEEtitleabstractindextext{%
\fcolorbox{abstractbg}{abstractbg}{%
\begin{minipage}{\textwidth}%
\begin{wrapfigure}[17]{r}{4in}%
\includegraphics[width=4in]{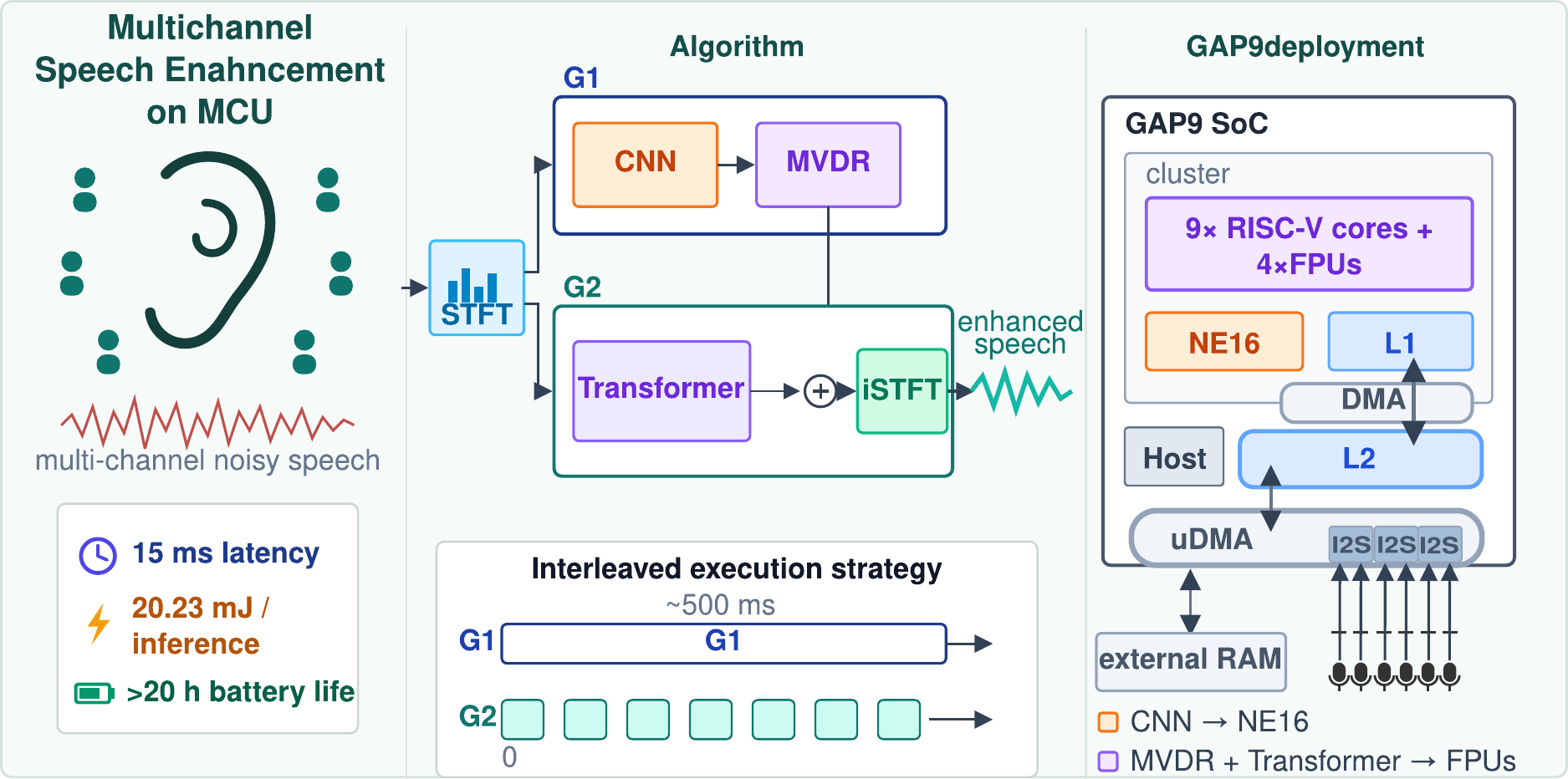}%
\end{wrapfigure}%
\begin{abstract}
Accurate, efficient, and low-latency spatial beamforming is a key
component in emerging smart hearable devices, enhancing speech while
suppressing noise and interference. However, handling multiple input
sources under strict real-time constraints pose significant
challenges for the low-power, resource-constrained microcontroller
units (MCUs) used in hearables. We present an optimized methodology
for the real-time execution of a neural-network-based minimum variance
distortionless response (MVDR) beamformer on MCUs. Using six
microphones and a three-stage mixed-precision scheme (\texttt{float32}
MVDR, \texttt{int8} CNN, \texttt{float16} Transformer), the pipeline
pairs a CNN that estimates the MVDR weights with a lightweight
Transformer that applies a per-frame correction. By time-slicing
weight estimation with beamforming, it achieves a 15~ms per-frame
latency while refreshing a complete set of CNN-derived weights every
564~ms. The deployed mixed-precision pipeline attains a
short-time objective intelligibility (STOI) of 97.65\%, a
scale-invariant signal-to-noise ratio (SI-SNR) of 20.26~dB, and a
wideband PESQ of 3.676 
at an average power of 45.9~mW. A speech activity detection (SAD)
module (98.5\% accuracy, 0.62~mJ per inference) bypasses the pipeline
during silence; under realistic deployment conditions, the system
exceeds the 16~h all-day target on a 100~mAh battery, with an
estimated lifetime of up to $\sim$20~h. To our knowledge, this is the
first real-time multi-channel Transformer-based neural beamforming
pipeline deployed on an MCU-class device.

\end{abstract}

\begin{IEEEkeywords}
beamforming, speech enhancement, MCU, real-time, Transformers
\end{IEEEkeywords}
\end{minipage}
}}

\maketitle

\section{Introduction}
Speech enhancement (SE) is a critical component in hearable devices, as it not only improves user communication by attenuating environmental noise but also enhances the reliability of interactions with voice-activated AI assistants.
Most existing approaches operate on single-channel inputs, since multi-channel processing imposes substantially higher computational and memory demands~\cite{single_vs_multi,single_vs_multi2,HearingAids,platform1}.
Despite this cost, multi-channel SE methods that use a microphone array, such as neural beamforming~\cite{NeuralBeamformingDefinition}, offer clear advantages: they suppress background noise more effectively and, importantly, preserve the spatial cues of the acoustic scene, maintaining the directionality of the target signal~\cite{whytouse}.
These properties are particularly relevant for hearables, where intelligibility and localization are central to the user experience.

At the same time, hearable devices are constrained by limited size and battery capacity, as they must fit unobtrusively inside or behind the ear.
These devices can only accommodate a small battery, which in turn caps the power available to the processing electronics and limits all-day operation and real-time performance.
In practice, sustaining several hours of continuous operation on a hearing-aid-class battery limits the power budget to the order of 100 mW~\cite{GAP9_HearDevice}.
Such constraints can only be met by microcontroller-class processors, which usually offer just a few MB of on-chip memory and a few billion multiply-accumulate (MAC) operations per second (GMAC/s), demanding careful co-design of the algorithm and its hardware mapping.

This work extends our previous single-branch neural beamformer~\cite{previous}, in which a single convolutional neural network (CNN) estimated the MVDR weights, by introducing a hybrid architecture that separates the estimation of the slowly varying acoustic scene from fast, per-frame adaptation.
A CNN estimates spectral masks used to compute minimum variance distortionless response (MVDR) beamforming weights, complex-valued per-frequency coefficients that, applied to the multi-channel short-time Fourier transform (STFT) of the microphone signals, perform spatial filtering to suppress noise while preserving the target direction, and a lightweight Transformer~\cite{Attention_is_all_you_need} produces an additive per-frame correction to those weights. 
Because the CNN captures the near-stationary structure of the audio scene, its output can be held over longer intervals; this lets the most demanding block of our pipeline, the CNN itself, run at a lower frequency, thus achieving higher energy efficiency, with no loss of quality.

The pipeline runs in a three-stage mixed-precision scheme: the MVDR stage is retained in \texttt{float32} for numerical stability, the CNN is quantized to \texttt{int8} and offloaded to the GAP9 NE16 accelerator, and the Transformer exploits the platform's native \texttt{float16} SIMD instructions. To meet the strict latency budget of our real-time constraint (15 ms), we interleave execution: the lightweight Transformer correction and beamforming run every frame to produce enhanced output in real time, while the heavier CNN weight estimation is amortized across many frames. The Transformer thus supplies fine-grained per-frame adaptability within a slower CNN update cycle. We further integrate an efficient speech activity detection (SAD) module to suppress computation during silence.
We evaluate the pipeline on GAP9, an ultra-low-power RISC-V SoC, across both algorithmic accuracy and hardware deployment metrics, demonstrating real-time operation while preserving enhancement quality and extending advanced multi-channel SE to wearable microcontroller-class devices.
\noindent Our main contributions are summarized as follows:

\begin{itemize}

\item We propose a hybrid CNN–Transformer neural beamforming architecture in which a lightweight causal Transformer produces an additive per-frame correction to MVDR beamforming weights from raw six-channel STFT input, extending our prior single-CNN beamformer with inference-time adaptability and a separation of temporal scales that improves both accuracy and deployment efficiency.

\item We deploy the pipeline under a three-stage mixed-precision scheme: a \texttt{int8} CNN on the NE16 accelerator, a \texttt{float16} Transformer exploiting the SoC's native SIMD, and a \texttt{float32} MVDR for numerical stability.

\item We deploy the complete pipeline on the GAP9 ultra-low-power SoC through hardware-aware optimization beyond the standard toolflow, achieving real-time operation with all-day autonomy on a 100~mAh Li-Po battery.

\end{itemize}
On the VoiceBank dataset~\cite{VoiceBank}, the proposed system achieves a STOI of 98.01\% and a PESQ-WB of 3.768 at 24.2~mJ per inference. On a 100~mAh, 3.7~V battery representative of a behind-the-ear hearable, assuming speech activity 30\% of the time~\cite{smeds_selecting_2020} and including six 0.9~mW MEMS microphones as a conservative upper bound, the system sustains more than 20~h of continuous operation. To our knowledge, this is the first real-time multi-channel Transformer-based neural beamforming pipeline deployed directly on an MCU-class device. code avaiable at~\url{https://github.com/Bomps4/Minimal_beamforming_Transformer}.

\section{Related Work}

Speech enhancement algorithms deployed on MCUs have predominantly relied on single-channel recurrent neural networks (RNNs). 
An early example is TinyLSTM~\cite{TinyLSTMs}, deployed on the STM32F746VE microcontroller, where the authors start from baseline LSTM-based SE architectures and progressively compress them through structured pruning of hidden units, skipping RNN state updates, and 8-bit quantization of weights and activations. Employing training-aware quantization to reduce the degradation introduced by quantization.
Another representative solution is given by the NNoM framework~\cite{NNOM}, which demonstrates the deployment of the RNNoise~\cite{RNNNoise} model on a single-core ARM Cortex-M MCU.
Owing to its compact GRU layers, roughly 87k parameters, and carefully constrained activation ranges, RNNoise can be effectively quantized to 8 bits with minimal loss in quality.
Both of these approaches, however, are limited to relatively simple mono-core MCUs.
In contrast, more recent work has explored multi-core MCUs such as GAP9. For instance, Rusci \textit{et al.}~\cite{Rusci} introduced a mixed-precision deployment scheme that quantizes recurrent layer parameters and activations to \texttt{int8}, while retaining \texttt{float16} precision for other tensors. 
This strategy achieves an almost lossless deployment, with only a 0.007 degradation in STOI.

Despite these advances, all of the above methods remain restricted to a single-channel SE, and none address the more challenging problem of deploying multi-channel, beamforming-based SE pipelines under MCU-class resource constraints.
More recently, several neural architectures have been proposed for multichannel SE~\cite{Lightweight,Kovalyov2023DFSNet,lee2024deftan,UDenoise}. 
These approaches span a broad design space, from UNet-like encoder–decoder structures~\cite{UDenoise} to Transformer-based models~\cite{Lightweight,Kovalyov2023DFSNet,lee2024deftan}, and demonstrate strong denoising capabilities. 
However, despite their promising accuracy and, in some cases, computational requirements comparable to those of our pipeline ~\cite{Lightweight}, none of these methods have yet been deployed on real hardware platforms. 
Conversely, in this paper we present an end-to-end deployed solution. This brings multichannel SE with beamforming into an MCU-class device under realistic deployment constraints. 

Finally, we also report the work of Wang\textit{et al.}\cite{MVDRALtri}. They do not use the MVDR beamformer for speech enhancement, but rather use it to estimate the position of a sound source. They report taking 1.5 seconds to run the MVDR algorithm (6.5$\times$ slower than our weight estimation step) falling short of real-time on a 200 MHz PIC32MZ microcontroller.

\section{Methods}
In the following paragraphs, we describe the beamforming algorithm, according to the dataflow shown in Fig.~\ref{fig::structure}, from the multichannel audio input to the enhanced output.
After conversion to the time-frequency domain, the pipeline is divided into two main stages.
G1 estimates the beamforming weights and consists, following the implementation of Ceolini et al.~\cite{Combine_}, of a neural mask-estimation stage (G1\_1) followed by MVDR weight computation (G1\_2).
G2 then applies the estimated weights to produce the enhanced signal.
We augment this formulation by adding a Transformer network to directly predict a correction to the MVDR weights.  

\subsection{Signal model and CNN neural beamforming}
\label{sec:Sign_model_CNN}
\begin{figure}
    \includegraphics[width=\columnwidth,keepaspectratio]{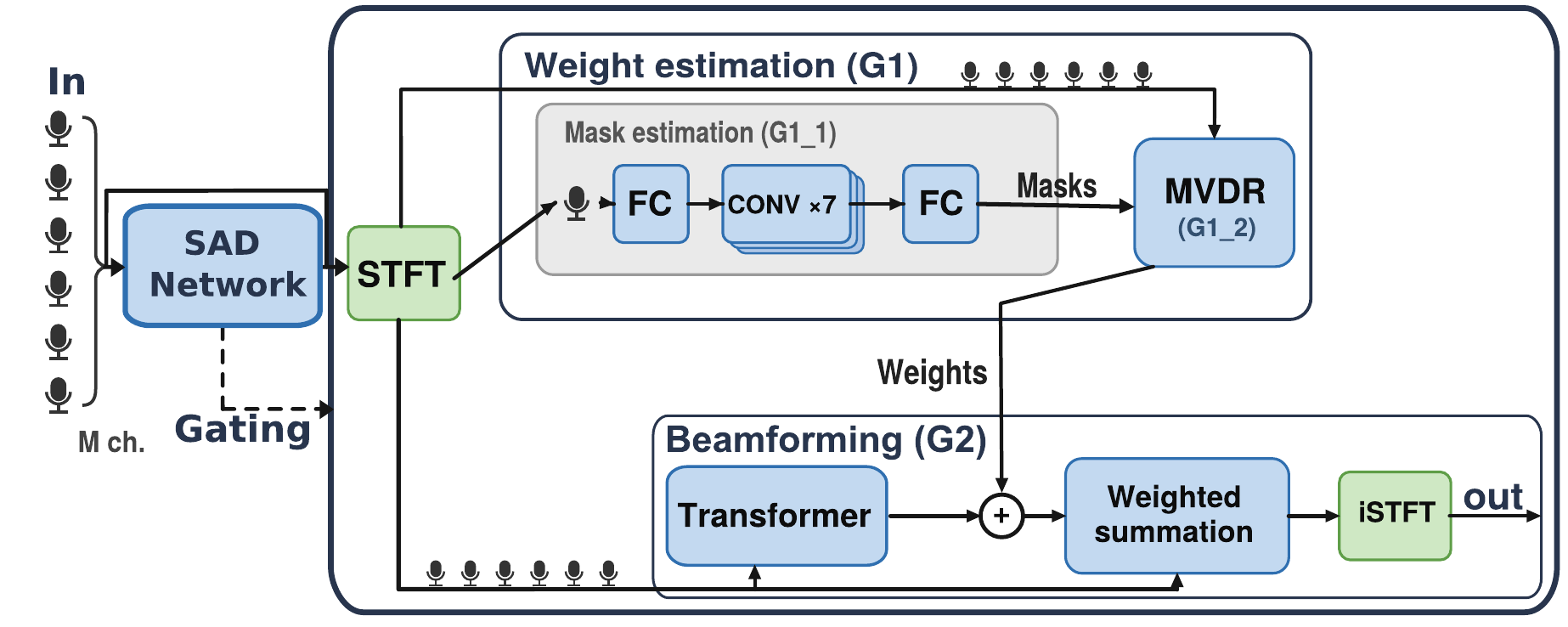}
    
    \caption{Block diagram view of the elements in our neural beamforming application.}
    \label{fig::structure}
\end{figure}

We consider a single target speech source captured by $M$ microphones in the presence of noise. Following the signal model adopted by Ceolini et al.~\cite{Combine_}, the signal at microphone $m$ is modeled as
\begin{equation}
    y_m(t) = s(t) * r_m(t) + k_m(t),
\end{equation}
where $s(t)$ is the clean speech signal, $r_m(t)$ is the room impulse response, $k_m(t)$ is the additive noise, and $*$ denotes convolution.

Under the narrowband approximation~\cite{why_STFT}, applying the short-time Fourier transform (STFT)~\cite{sourcespeech} gives
\begin{equation}
    Y_m(\omega,n)
    =
    S(\omega,n)R_m(\omega,n) + K_m(\omega,n),
\end{equation}
where $\omega$ denotes the frequency bin and $n$ the frame index. In the following, we denote by $t_{\mathrm{stft}}$ the amount of audio covered by each STFT window and by $t_{\mathrm{hop}}$ the stride between consecutive windows, both expressed in milliseconds.

We perform speech enhancement using a minimum-variance distortionless response (MVDR) beamformer, which minimizes interference while preserving the target signal.
Rather than estimating an explicit steering vector, the adopted formulation derives the beamforming weights from the spatial covariance matrices of speech and noise~\cite{ImprovedMVDR}. 
Since the speech and noise components are not observed separately, these covariance matrices cannot be computed directly.
G1, therefore, first employs a neural network to estimate speech and noise masks from the STFT of a reference microphone (G1\_1).
The two masks are then used together with the multichannel STFT observations to estimate the speech and noise covariance matrices and compute the MVDR weights (G1\_2).

For mask estimation, we employ the compact CNN referred to as MUDV4 C4 512 in~\cite{Combine_}. The network operates on the STFT representation of a single reference microphone, G1\_1 in Figure~\ref{fig::structure}.
Since the STFT coefficients are complex-valued, each coefficient is decomposed into its real and imaginary components, which are stacked as two real-valued channels.
This Cartesian representation is a linear bijection between $\mathbb{C}$ and $\mathbb{R}^{2}$ and, unlike a magnitude-phase representation, does not introduce a nonlinear coordinate transformation.

The CNN, whose design we take from ~\cite{Combine_}, consists of a fully connected layer that expands the input dimensionality, followed by a stack of dilated 2-D convolutions with residual connections, PReLU activations, and normalization layers. A final projection produces a single-channel speech mask
\begin{equation}
    M_{ss} \in \mathbb{R}^{F \times N},
\end{equation}
where $F$ is the number of frequency bins and $N$ the number of STFT frames processed by the network. The corresponding noise mask is obtained as
\begin{equation}
    M_{nn} = 1 - M_{ss}.
\end{equation}
Avoiding a separate output head for the noise mask, together with the small number of convolutional features, limits the network to approximately 20\,k parameters.
Nevertheless, because the network operates on one second of audio, a complete inference requires approximately 342\,MMACs.

The masks estimated by G1\_1 are combined with the multichannel STFT observations to estimate the spatial covariance matrices for speech and noise.
Denoting the multichannel observation at frequency $\omega$ and frame $n$ by
\begin{equation}
    \mathbf{Y}(\omega,n)
    =
    [Y_1(\omega,n),\ldots,Y_M(\omega,n)]^{T},
\end{equation}
the covariance matrices are computed as
\begin{equation}
    \mathbf{\Phi}_{vv}(\omega)
    =
    \sum_{n=1}^{N}
    M_{vv}(\omega,n)
    \mathbf{Y}(\omega,n)\mathbf{Y}^{H}(\omega,n),
    \qquad vv \in \{ss,nn\},
\end{equation}
where $(\cdot)^H$ denotes the Hermitian transpose.

G1\_2 then computes the minimum-variance distortionless response (MVDR) beamforming weights. The MVDR beamformer minimizes interference energy while preserving the target signal.
Following the formulation of~\cite{ImprovedMVDR} we get
\begin{equation}
    \mathbf{w}_{\mathrm{MVDR}}
    =
    \frac{
        \mathbf{\Phi}_{nn}^{-1}\mathbf{\Phi}_{ss}
    }{
        \operatorname{tr}
        \left(
            \mathbf{\Phi}_{nn}^{-1}\mathbf{\Phi}_{ss}
        \right)
    },
    \label{eq::weight}
\end{equation}
where $\mathbf{\Phi}_{nn}$ and $\mathbf{\Phi}_{ss}$ denote the noise and speech spatial covariance matrices, respectively, and $\operatorname{tr}(\cdot)$ denotes the trace.
This completes the weight estimation performed by G1.

\subsection{Transformer-based beamforming weight correction}
To improve the adaptability of the estimated beamforming weights at inference time, we introduce a lightweight Transformer network in the Beamforming block (G2) that produces a causal, additive correction to the MVDR weights.
The correction is estimated directly from the raw multi-channel STFT input, without access to the CNN's internal representations or the MVDR-computed weights.
The input to the correction network is formed by stacking the STFT of the current and previous frames across all M microphones. 
Each frame is decomposed into real and imaginary parts, yielding an input tensor of shape F$\times$D, where F is the number of frequency bins and D$= 2 \cdot 2 \cdot M$ is the model dimension, accounting for two frames, two components (real and imaginary), and M microphones.
In our case M$ = 6$, giving D$=24$.

The network consists of two stages.
First, a 1D convolutional layer applied along the frequency axis processes the input tensor, extracting local frequency-domain patterns while preserving the model dimension D.
Second the output is then fed into a single Multi-head Attention encoder layer with H$=$ 2M $=$ 12 attention heads, treating the frequency bins as the sequence dimension. Each attention head therefore operates on a subspace of dimension ${\text{D}}/{\text{H}}=2$, corresponding to one real-imaginary pair per microphone per frame. The outputs of all heads are concatenated, recovering the full model dimension D, and passed through a linear projection layer that maps the representation to a complex-valued correction tensor of shape F $\times$ M, matching the shape of the MVDR beamforming weights.
The correction is added directly to the MVDR weights prior to their application to the STFT frames:
\begin{equation}
w_{\text{final}}(\omega) = w_{\text{MVDR}}(\omega) + \Delta w(\omega),
\end{equation}
where $w_{\text{MVDR}}$($\omega$)$ \in \mathbb{C}^{M}$ are the weights produced by the CNN-MVDR pipeline and $\Delta w$($\omega$) $\in \mathbb{C}^{M}$ is the correction output by the Transformer. The corrected weights $w_{\text{final}}$($\omega$) are then normalized and applied to the multi-channel STFT as described in the previous section.

\subsection{Speech Activity Detection (SAD) network}

In the absence of speech, estimating speech–noise masks is an ill-defined task; to avoid performing it, we introduce a lightweight SAD network that skips mask estimation and beamforming entirely when no speech is detected.
The network is a 1-D ResNet-8~\cite{he2016residual} operating on the raw waveform prior to the STFT, composed of an initial convolution, three residual blocks (the first without a pointwise convolution in the skip path), ReLU activations, a global max-pooling layer, and a fully connected head yielding a binary speech/non-speech decision.
Although it has 7.6k more parameters than the G1 CNN network (27.6k in total), it is far cheaper to run, requiring $6\times$ fewer operations than the full speech-enhancement pipeline (76 MMACs), and its structure has been tuned to be efficiently executed on our platform, thereby avoiding substantial latency increase during normal operation. 

\subsection{Interleaved execution strategy}
\label{sec:int_strat}
\begin{figure}
    \includegraphics[width=\columnwidth]{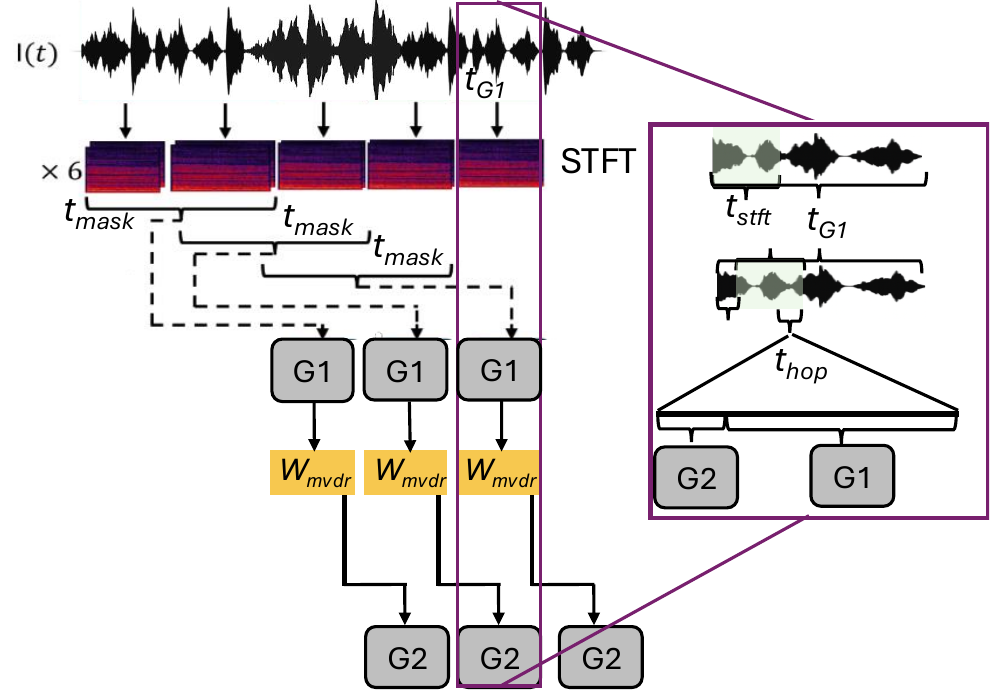}
    
    \caption{Execution schedule of our proposed neural beamforming pipeline.}
    \label{fig::schedule}
\end{figure}
The processing pipeline described so far comprises speech activity detection (SAD), STFT computation, MVDR weight estimation, Transformer-based weight correction, beamforming, and inverse STFT. If executed sequentially at every processing pass, the full pipeline would require 443.6 million MACs. Of this total, the SAD network accounts for 76 MMACs, while the speech-enhancement path accounts for the remaining 367.6 MMACs.

Within the speech-enhancement path, the STFT is first computed over the six input channels, requiring 7.7 MMACs. The resulting multi-channel spectral representation is then consumed by two subsequent processing stages. The first stage, denoted as G1, estimates the MVDR beamforming weights. It consists of the CNN-based mask estimator, G1\_1, which requires 342 MMACs, followed by the MVDR weight computation, G1\_2, which requires an additional 12 MMACs. Overall, G1 requires 354 MMACs, making it by far the dominant computational component of the pipeline.

The second stage, denoted as G2, performs the operations required to produce the enhanced output for each frame. In contrast to G1, G2 is comparatively lightweight, requiring only 5.9 MMACs. Of these, 5.7 MMACs are due to the Transformer-based weight correction, while the remaining approximately 0.2 MMACs are due to the weighted summation and inverse STFT.

The higher computational requirements of G1 motivate the introduction of our \textit{interleaved execution strategy}.
As illustrated in Figure~\ref{fig::schedule}, execution begins by buffering the required number of audio samples ($\textit{t}_\textit{mask}$) to compute the initial set of beamforming weights.
Once the first set of weights has been obtained, we call the time necessary for this computation $\textit{t}_\textit{G1}$, the system transitions into the interleaved regime: at each hop ($\textit{t}_\textit{hop}$), the current weights along with the transfomer produced correction are applied to perform beamforming along with the inverse STFT (G2), while the remaining computational budget within the same hop is used to advance G1.
In this manner, G1 progresses incrementally, typically one neural network layer at a time, until the next set of weights is ready.
Crucially, although the full computation of new weights requires $\textit{t}_\textit{G1}$, the execution of the G2 continues without interruption during this period, ensuring that enhanced outputs are generated in real time while weight estimation proceeds concurrently in the background.

\subsubsection{Hardware platform \& deployment}
\label{HardwareDeploymentCons}
\begin{figure}[h]
    \centering
    \includegraphics[width=0.8\columnwidth]{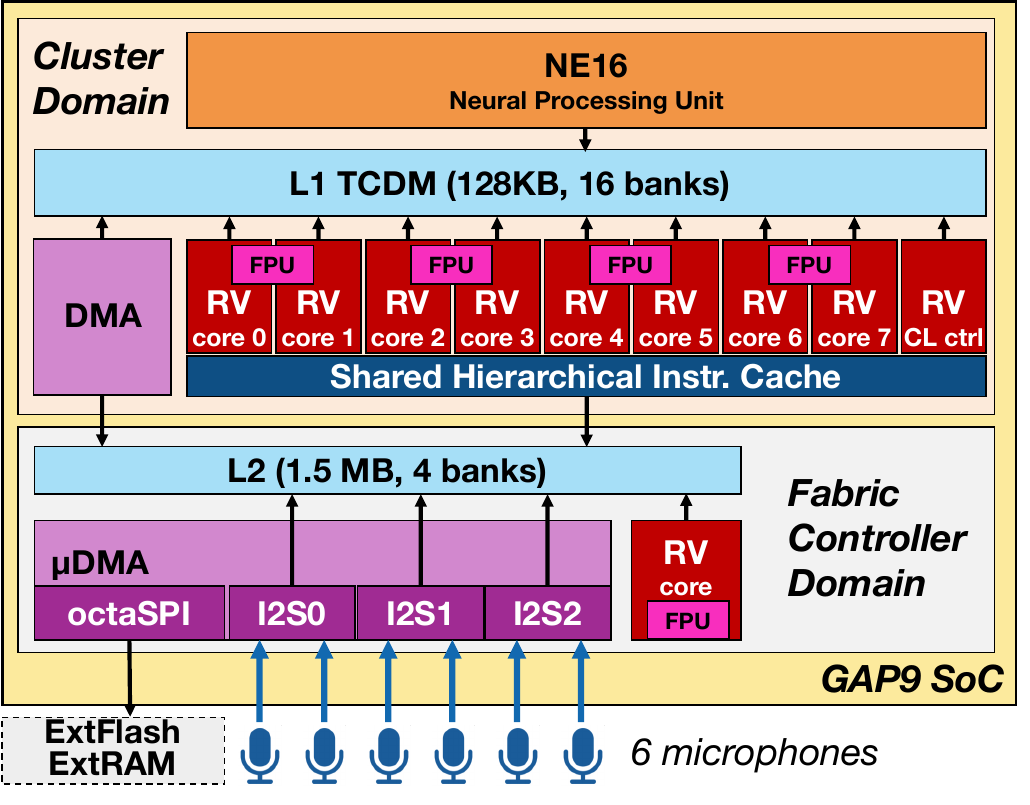}
    \caption{Simplified architecture of the GAP9 SoC in the investigated configuration.}
    \label{fig:GAP9Struct}
\end{figure}

We deploy our neural beamforming application on the GAP9 SoC from GreenWaves Technologies, see Figure~\ref{fig:GAP9Struct}. GAP9 is a low-power heterogeneous platform that combines a RISC-V host (FC) core with 1.5~MB of scratchpad L2 memory and a cluster (CL) of 9 RISC-V cores, supported by the NE16 neural engine accelerator and 4 floating-point units (FPUs). The host and cluster cores run at up to 370~MHz at 0.80~V; both the supply voltage and the clock frequency are configurable in software, with an energy-optimal operating point at 240~MHz and  0.65~V. The cores feature vector MAC instructions capable of performing 4$\times$8-bit dot products per cycle, while the NE16 integrates 9$ \times$9 $\times$16 8$\times$ 1-bit MAC units, sustaining up to 150 8-bit MAC operations per cycle,  about 5 $\times$ faster than the cores alone. The cluster also provides a 128~kB single-cycle L1 scratchpad memory, complemented by DMA engines that efficiently manage transfers to the larger L2 memory and to external memories (32~MB RAM and 64~MB FLASH connected via octaSPI).

We deploy our networks on the device using the GapFlow toolchain. Since GapFlow does not provide optimized kernels for the attention mechanism of Transformer layers, we implemented a custom attention kernel, tailored to the cluster architecture and exploiting the \texttt{float16} SIMD capabilities of the FPUs, to efficiently execute the weight-correction Transformer within the real-time budget.

As the beamforming pipeline requires six input spectrograms to run, storing all of them in L2 would consume 800~kB of memory, impeding the execution of G1\_1. Since the six spectrograms are only needed from the MVDR step onward, we keep them in the larger external RAM. To maximize the efficiency of our most critical component, G2, we retain a buffer in the faster L2 memory that holds data for each of the six microphones, each storing up to $2\times\textit{t}_\textit{stft}$ of audio for the G2 block.

\begin{figure}[h]
    \centering
    \includegraphics[width=0.5\linewidth]{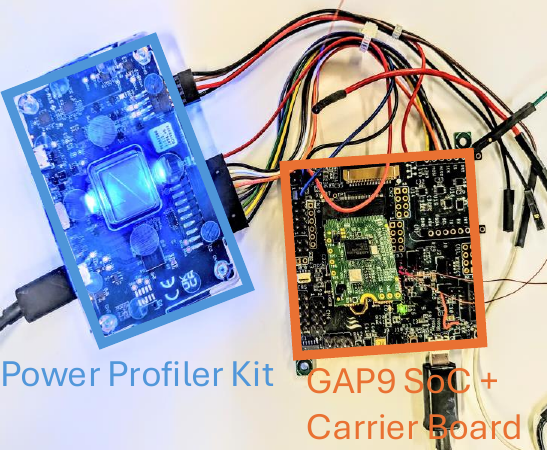}
    \caption{Experimental setup }
    \label{fig:sperimentale_apparato}
\end{figure}

\section{Experimental Results}
We evaluate the proposed neural beamforming pipeline on GAP9 across memory, quantization, performance, and energy efficiency, and compare it with state-of-the-art speech enhancement systems deployed on an MCU.
{
\subsection{Setup \& model training}}

We evaluate our approach using a simulated multi-microphone scenario of a single speaker in a reverberant environment with diffuse noise, which we simulate using gpurir~\cite{gpuRIR}. 
For beamforming experiments, following the work of~\cite{Slow_Fast}, we adopt the Voice Bank + DEMAND (VBD) dataset~\cite{VBD}. The training set is generated by mixing clean speech from 28 speakers of the Voice Bank corpus~\cite{VoiceBank} with 10 types of noise from the DEMAND corpus~\cite{DEMAND} at four signal-to-noise ratio (SNR) levels (0, 5, 10, and 15 dB). 

The test set uses the remaining 2 speakers and 5 noise types unseen during training, mixed at SNR values of 2.5, 7.5, 12.5, and 17.5 dB. 
All signals are sampled at 8~kHz following ~\cite{Combine_}. For each sample, we select a room size whose dimension are defined by uniforms distribution over width, height and depth $U(3,6) \times U(3,6) \times U(3,6)$ meters, a reverberation time $T_{60} \sim U(0.2,0.8)$ seconds, one noise recording, and the positions of six microphones and the source, without imposing constraints on their relative distances.
Following the original implementation~\cite{Combine_}, the network is trained using the scale-invariant signal-to-distortion ratio between the undistorted signal and the one reconstructed by the beamforming pipeline. 
We use the Adam optimizer, a learning rate of $10^{-4}$, and early stopping with a patience of 2 epochs for regularization. 
The capabilities of our pipeline are measured using the STOI~\cite{STOI} and ESTOI~\cite{ESTOI} metric for intelligibility, while we use SI-SNR~\cite{SISNR} and PESQ-wideband~\cite{PESQ} to measure audio quality.

For the SAD experiments, we start from the 121 noise types available by combining the NoiseX-92~\cite{Noisex92} dataset and DEMAND.
To further enrich the variability of background conditions, we augment the noise by introducing composite noise types that mix pairs of noises drawn from the two corpora, leading, in principle, to up to $121 \times 60$ possible combinations.
Together with the Voice Bank speech corpus, this yields a dataset with the following statistics: training set of 15,034 samples (40.14\% noise, 59.86\% speech+noise), validation set of 1,726 samples (42.06\% noise, 57.94\% speech+noise), and test set of 1,300 samples (38.46\% noise, 61.54\% speech+noise).
All signals are sampled at 8~kHz to be consistent with the beamforming pipeline.
To train the SAD, we use the Adam optimizer with a learning rate of $10^{-4}$ and a binary cross-entropy loss.
We evaluate the results on the test set, reporting accuracy, precision, and recall; the latter is particularly important, as a false negative means that the current speech would be completely ignored by the entire pipeline. 

Following the work of Ceolini~\cite{Combine_}, all experiments use an STFT window length of $\textit{t}_\textit{stft}=64$~ms and hop size of $\textit{t}_\textit{hop}=15$~ms.
Each input sequence spans 1~s of audio, corresponding to the total duration processed by the network ($\textit{t}_\textit{mask}$).
For deployment on the GAP9 platform, buffer sizes are determined directly from these time windows and the signal's data rate. The energy measurement have been performed using the Nordic Power Profiler Kit, see Figure~\ref{fig:sperimentale_apparato} for our experimental setup.

\subsection{Baselines}
\label{sec:baseline}

\begin{table}[t]
\centering
\caption{Speech enhancement performance of the CNN, Transformer, and hybrid
Transformer\,+\,CNN architectures, with the unprocessed noisy input as reference.
Values are mean\,$\pm$\,std.}
\label{tab:arch_comparison}
\begin{tabular}{lcccc}
\toprule
Model & STOI\,(\%) & ESTOI\,(\%) & PESQ-WB & SI-SNR\,(dB) \\
\midrule
Noisy               & 92.38\,{\scriptsize$\pm$0.28} & 86.97\,{\scriptsize$\pm$0.33} & 2.609\,{\scriptsize$\pm$0.034} & 10.15\,{\scriptsize$\pm$0.13} \\
\midrule
CNN                 & 94.19\,{\scriptsize$\pm$0.23} & 89.95\,{\scriptsize$\pm$0.47} & 2.947\,{\scriptsize$\pm$0.041} & 17.70\,{\scriptsize$\pm$0.31} \\
Transformer         & 95.98\,{\scriptsize$\pm$0.12} & 91.96\,{\scriptsize$\pm$0.17} & 3.262\,{\scriptsize$\pm$0.013} & 19.79\,{\scriptsize$\pm$0.10} \\
\makecell[l]{CNN + \\Transformer}            & \textbf{98.04}\,{\scriptsize$\pm$0.07} & \textbf{95.52}\,{\scriptsize$\pm$0.12} & \textbf{3.747}\,{\scriptsize$\pm$0.018} & \textbf{22.02}\,{\scriptsize$\pm$0.12} \\
\bottomrule
\end{tabular}
\end{table}

Table~\ref{tab:arch_comparison} evaluates our complete pipeline against the unprocessed noisy input, which serves as reference. The pipeline raises SI-SNR from 10.15 to 22.02~dB (+11.87~dB), PESQ-WB from 2.61 to 3.75 (+1.14), STOI from 92.38 to 98.04, and ESTOI from 86.97 to 95.52.

To isolate the contribution of each stage, we evaluate the CNN and Transformer independently.
In isolation, the CNN operates as in the full pipeline, estimating the masks which are used to derive the MVDR weights, whereas the Transformer predicts the beamforming weights directly; in the combined pipeline, it instead predicts a per-frame correction to the MVDR weights.

As shown in Table~\ref{tab:arch_comparison}, their combination improves all metrics: +2.23~dB SI-SNR, +3.56 ESTOI, +2.06 STOI, and +0.49 PESQ-WB over the standalone Transformer, and +4.32~dB SI-SNR, +5.57 ESTOI, +3.85 STOI, and +0.80 PESQ-WB over the standalone CNN. 
Their composition requires no additional fusion or projection layers, so these gains are achieved without increasing model capacity.

The two stages operate on different temporal scales of the same STFT representation. The CNN, via stacked causal convolutions, integrates a receptive field of one second, capturing the slowly varying, near-stationary structure of the scene, noise statistics and spatial configuration.
The Transformer, instead, operates over a short ($\sim$30~ms) window, giving it the flexibility to track fast spectro-temporal details.

\subsection{Mixed-precision quantization}
\label{sec:mix_precision}



To reduce the memory footprint and computational cost of our deployed application, we convert the trained \texttt{float32} model into reduced-precision formats (\texttt{float16} or \texttt{int8}) using GAPflow's post-training quantization procedure, calibrating the quantization ranges on a randomly selected subset of the training data.
For on-device reproducibility, we fix the random noise realisation so that all configurations are evaluated on an identical input; under this fixed noise the noisy reference scores STOI 92.02\%, ESTOI 86.21\%, PESQ-WB 2.609, and SI-SNR 10.02 dB.
The fully \texttt{float32} model raises these to STOI 98.01\%, ESTOI 95.49\%, PESQ-WB 3.768, and SI-SNR 21.90 dB. Reduced precision affects the pipeline unevenly.
The neural components are robust: quantizing the CNN to \texttt{int8} incurs only marginal degradation (STOI 97.98\%, ESTOI 95.42\%, PESQ-WB 3.760, SI-SNR 21.54 dB), and quantizing the Transformer to \texttt{float16} produces no measurable change in any metric.

The MVDR beamforming stage, in contrast, is highly sensitive: it relies on covariance estimation and matrix inversion, which are numerically ill-conditioned~\cite{stab_MVDR}, and quantizing it to \texttt{int8} collapses the output entirely, while even \texttt{float16} degrades intelligibility below that of the noisy input.
We therefore retain the MVDR and beamforming stage (G1\_2 and G2 weighted summation) in \texttt{float32} in all configurations and apply reduced precision only to the neural components.

\subsection{Latency Constraint on Weight Estimation}
\label{sec:Lat_const}

\begin{figure}[b]
\includegraphics[width=\columnwidth]{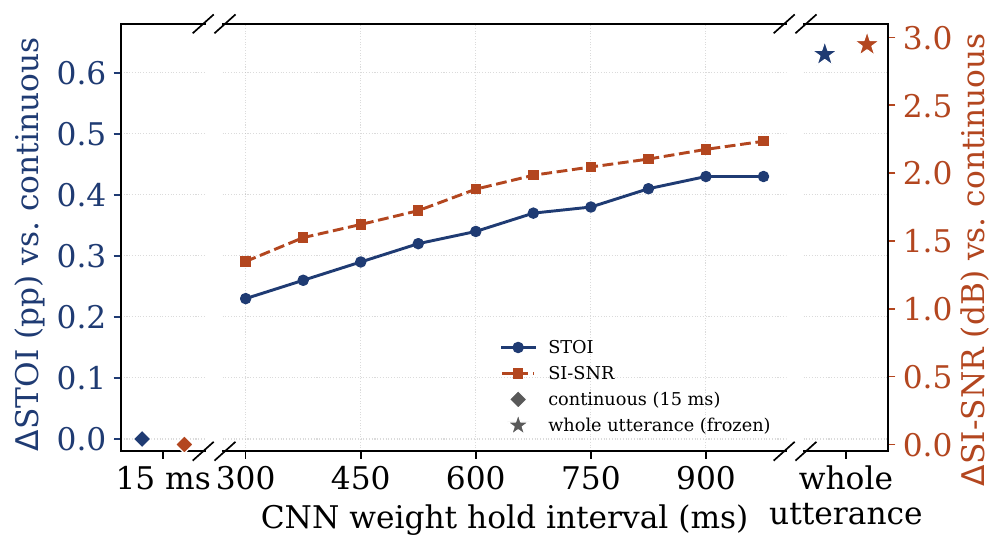}

\caption{Effect of the delay on the STOI and SI-SNR metrics. The extrema indicate the cases where there is no delay and where the CNN estimation is performed only once for the full utterance. }
\label{fig:delay}
\end{figure}

Before deploying the pipeline in the interleaved configuration described in Section~\ref{sec:int_strat}, we evaluate the system's tolerance to delays between weight estimation and beamforming.
As shown in Figure~\ref{fig:delay}, holding the CNN-derived weights fixed over a longer interval within a single utterance yields better enhancement than continuously re-chasing the estimate at every frame: quality improves monotonically as the hold interval lengthens, and a single estimate held for the whole utterance is optimal.
This tolerance, however, does not justify estimating the weights once and freezing them indefinitely. 
The per-utterance optimum reflects the within-utterance stationarity of the acoustic scene; across utterances the scene changes, a different speaker, a moved source, so the weights must be refreshed to track these transitions. The system must therefore continuously process the input stream and produce updated weights at a rate that keeps pace with the incoming audio. This imposes a latency constraint on G1: its effective weight-production interval must not exceed the span of audio it consumes per update, so that lag does not accumulate over a continuous stream.

\begin{figure*}[t]
    \includegraphics[width=\textwidth]{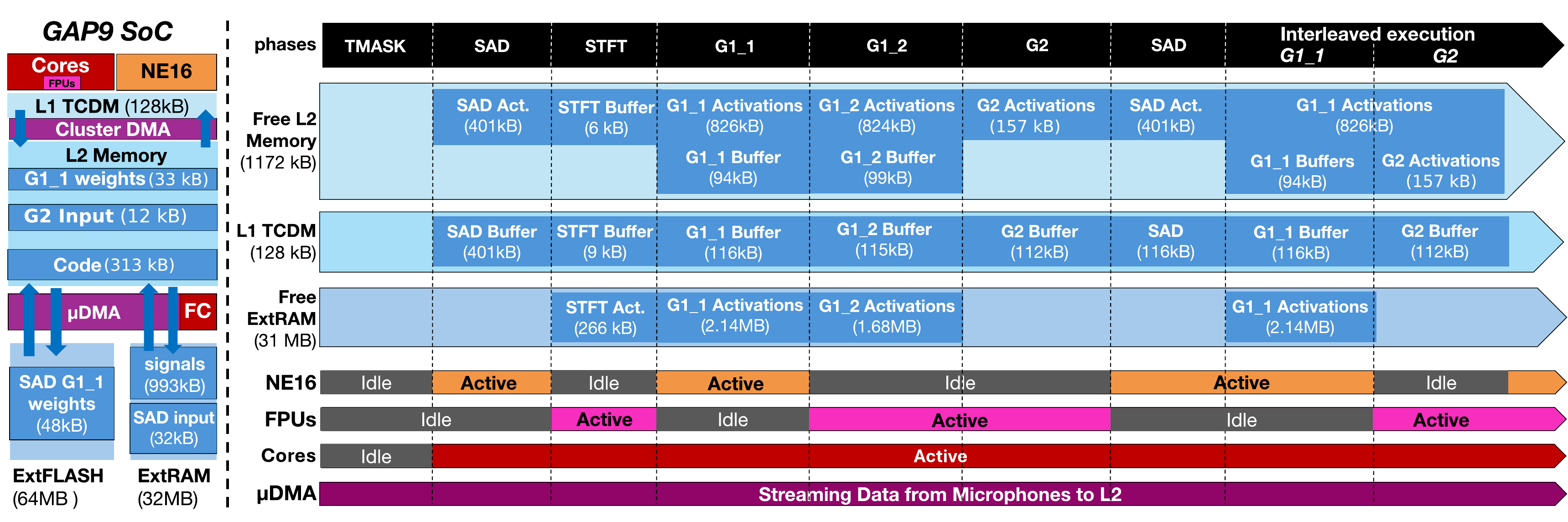}
    \caption{(left) Buffer memory allocation and data flow in GAP9 SoC.
(right) Beamforming pipeline resource allocation and scheduling in the various execution phases.}
    \label{fig:structure_and_pipeline}
\end{figure*}

\subsection[]{\textcolor{subsectioncolor}{Deployment on GAP9 SoC}}

As established in the previous section, each processing stage must complete within the time span of the input signal it consumes. In the case of the mask-estimation CNN, the input is one second of audio; therefore, its execution time must be less than one second to sustain continuous real-time operation. This requirement rules out floating-point deployment on the target platform. Even when operating at the maximum clock frequency of 370~MHz, the CNN alone requires approximately 1.1~s in \texttt{float16} to process one second of audio. 

Quantization to \texttt{int8} is therefore required for real-time operation. In this configuration, software execution on the general-purpose cores still requires 176.2~M cycles for the mask estimator, corresponding to 476~ms at the maximum 370~MHz operating frequency. Offloading the same computation to the NE16 accelerator, Figure ~\ref{fig:structure_and_pipeline}, reduces the cost to 74~M cycles, yielding a 2.38$\times$ reduction and lowering the measured mask-estimation runtime to 203~ms.

The second block, G2, must instead respect the application's per-frame latency constraint (15 ms), since it executes on every output frame: it applies the Transformer correction to the MVDR weights, performs the weighted summation, and computes the inverse STFT. 
\subsubsection{Transformer Deployment}
The Transformer deployment code generated by NNTool required approximately 9.25 million cycles per execution and a total of 1.1 MB of L2 memory together with 400 kB of off-chip L3 memory, exceeding both the execution-time and memory budgets of the target platform. The reliance on L3 memory, in particular, significantly increased execution time due to higher access latency than on-chip memories. The reliance on the external memory was due to the explicit materialization of the complete (F $\times$ F) attention-score matrix, where (F) is the number of frequency bins used as the sequence dimension ((F=257) in our case). We have implemented a series of optimizations to overcome the limitation of the NNtool's baseline implementation; a summary of the optimizations is reported in Table~\ref{tab}.

A first refinement targeted the attention mechanism's memory footprint by restructuring the computation across attention heads. Instead of materializing the full attention tensor for all heads simultaneously, the computation was serialized over the head dimension, such that only one (F $\times$ F) attention matrix is instantiated at a time.
This approach reduces peak memory usage compared to the baseline implementation and lowers the execution time to approximately 3.70 million cycles. However, it still requires an additional L2 buffer of (132, kB). Given that G1 already occupies most of the available memory, this configuration exceeds the GAP9 L2 memory budget.

To address this limitation, the attention computation was reorganized so that each core processes a single score row at a time, avoiding the need to materialize the entire matrix for any head. The query-key product, softmax, and score-value product are evaluated as matrix-vector operations. All intermediate scores fit within L1 memory, which serves as temporary working memory and requires no additional L2 allocation. This row-wise implementation reduced the execution time to approximately 2.22 million cycles while eliminating the need for an additional L2 buffer.

Further improvements were achieved by fusing the query-key product, softmax, and score-value product into a single FlashAttention-inspired kernel~\cite{flash_attention}. This fusion avoids intermediate loads and stores that would otherwise occur when executing the operations as separate kernels. The online (streaming) softmax used in standard FlashAttention was also evaluated; however, it proved less efficient in this configuration because the additional exponential evaluations increased pressure on the four available FPUs. The final fused implementation retains one score row in L1 and executes the complete Transformer in approximately 1.65 million cycles, bringing the full G2 to 1.85 million cycles

\begin{table}[t]
\centering
\caption{Transformer attention optimizations.}
\label{tab}
\begin{tabular}{@{}lcl@{}}
\toprule
\textbf{Implementation} & \textbf{Cycles} & \textbf{Attention buffers} \\
\midrule
\makecell[l]{NNTool baseline,\\ full attention matrix}
  & $9.25\,\mathrm{M}$ & \makecell[l]{942\,kB L2\\ + 400\,kB L3} \\
  \hline
\makecell[l]{Sequential heads,\\ one full matrix per head}
  & $3.70\,\mathrm{M}$ & 132\,kB L2 \\
  \hline
Row-wise attention in L1
  & $2.22\,\mathrm{M}$ & L1 only \\
  \hline
Fused row-wise attention kernel
  & $1.65\,\mathrm{M}$ & L1 only \\
\bottomrule
\end{tabular}
\end{table}

\subsubsection{Full application Deployment}

\begin{table}[t]
\centering

\caption{Timing and energy comsumption on GAP9 under interleaved execution. We report for G2 the total energy consumed by all executions in the time it takes for a G1's inference to be completed. }
\label{tab:timing_energy}
\resizebox{\columnwidth}{!}{
\begin{tabular}{@{}lcc@{}}
\toprule
 & Base & Optimized \\
 & \makecell[c]{(G1 @ 0.8\,V, 370\,MHz)\\(G2 @ 0.8\,V, 370\,MH)}& \makecell[c]{(G1 @ 0.65\,V, 240\,MHz)\\(G2 @ 0.8\,V, 370\,MH)} \\
\midrule
G1 execution time [ms]        & 244 & 379 \\
\quad STFT (\texttt{fp32}, 2.4\,M cyc.) & 6.5 & 10 \\
\quad Mask est.\ (\texttt{int8}, NE16)  & 203 & 315 \\
\quad MVDR (\texttt{fp32}, 13\,M cyc.)  & 35  & 54 \\
\makecell[l]{G2 execution time \\(\texttt{fp16,fp32}, 1.85\,M cyc.)} & 5   & 5 \\
Full pipeline [ms]      & 366 & 564 \\
G2 executions per inference       & 24 & 37 \\
\midrule
Total energy per inference: G1 [mJ]     & 13.2 & 8.95 \\
Total energy per inference: G2 [mJ]     & 10.98 & 16.9 \\
\midrule
Active power, G1 [mW]        & 54.1 & 23.8 \\
Active power, G2 [mW]        & 90.0 & 90.0 \\
Average system power [mW] & \textbf{66.1} & \textbf{45.9} \\
\bottomrule
\end{tabular}
}
\end{table}
We now report the memory, timing, and energy budget of the final implementation. As shown in Figure~\ref{fig:structure_and_pipeline}, G1
is the most memory-hungry stage, occupying about 3.14~MB of L3 (including the multichannel audio and its STFT) and about 920~kB of
L2 during execution. Adding the resident program code (313~kB), the
G2 input buffers kept in L2 for fast access (24~kB), the 
G1\_1 and Transformer weights (33~kB), and the Transformer working
memory (157~kB) yields a peak L2 occupation of 1447~kB, within the
1.5~MB L2 of GAP9.

Table~\ref{tab:timing_energy} summarizes the timing and energy of the
full pipeline deployed on GAP9. Execution follows a static time-sliced
schedule: each 15~ms frame allots 10~ms to the weight-estimation block
G1 and 5~ms to the enhancement block G2, so the per-frame latency is
15~ms by construction. With the whole SoC at 0.8~V and 370~MHz, G1
requires 244~ms of standalone compute, i.e., 24.4 frame slices,
yielding a new set of beamforming weights every 366~ms, while G2
processes every frame with the most recently available weights. Each
inference costs 24.2~mJ (13.2~mJ for G1 and 0.45~mJ for each of
the 24 interleaved G2 ); since G1 and G2 are
time-multiplexed rather than concurrent, the average power draw is
the duty-cycle-weighted mean of their active powers, 66.1~mW.

\begin{figure}[t]
\includegraphics[width=\columnwidth,height=6cm]{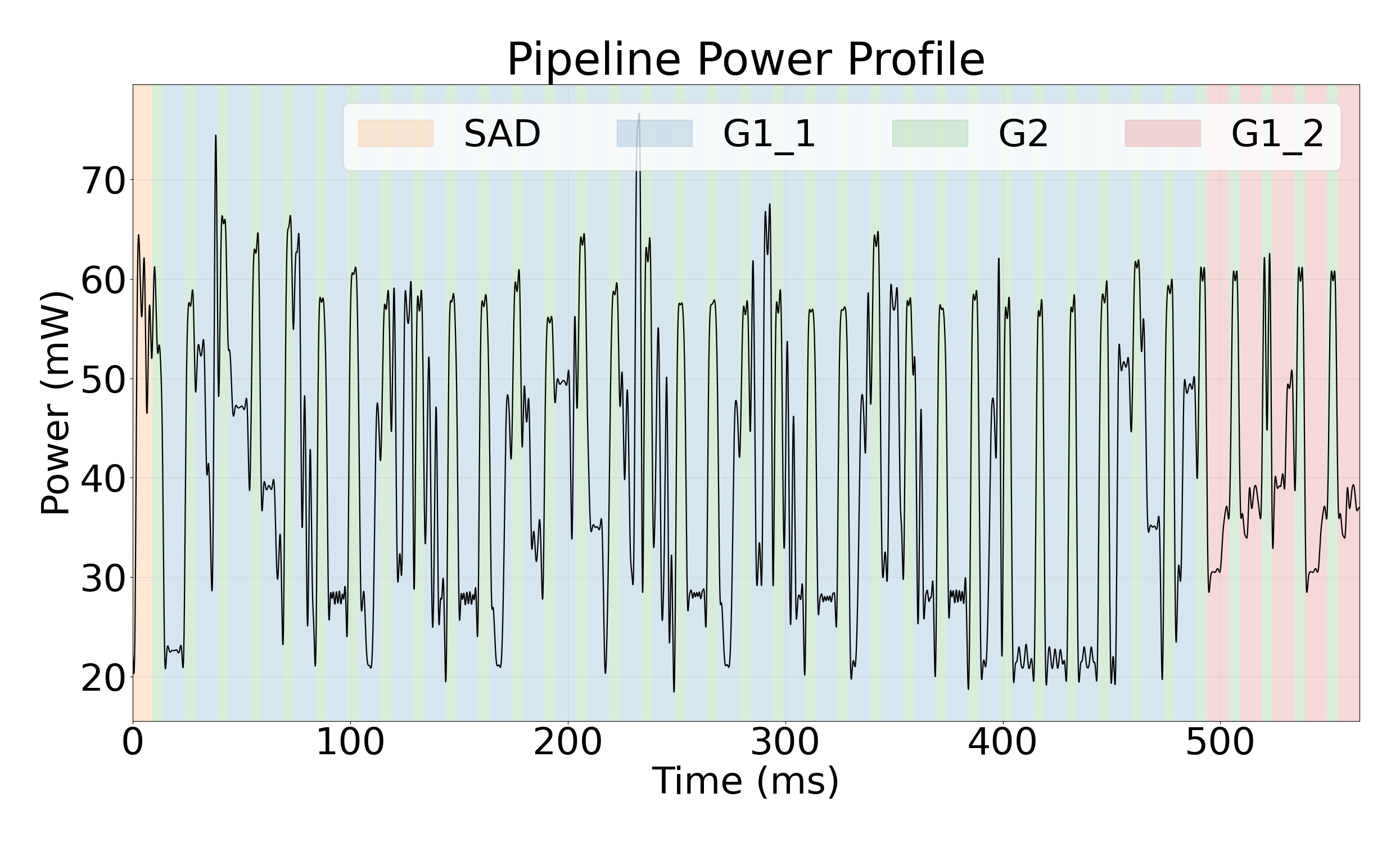}

\caption{Power profile of our application. }
\label{fig:profile}
\end{figure}

Our analysis, presented in Section~\ref{sec:Lat_const} showed, however, that slowing the mask-estimation CNN
does not degrade the speech enhancement quality: G1 can therefore be executed in a slower but more
efficient operating point (0.65~V, 240~MHz), stretching its execution time
time to 379~ms, moving the inference of the whole pipeline to 564~ms, this is shown in the power profile of our application Figure ~\ref{fig:profile}, but cutting G1's energy
per inference from 13.2 to 8.95~mJ.
The energy per inference of the full pipeline remains essentially unchanged
($\sim$25.9~mJ), but each inference now requires 564~ms, lowering
the average power draw to 45.9~mW, a 31\% reduction. This operating point is only viable thanks to the NE16 accelerator:
without it, G1 would require 796~ms of standalone compute at 240~MHz,
i.e., an effective inference period of $\sim$1.2~s under the interleaved
schedule, violating the 1-second constraint of
Section~\ref{sec:Lat_const}: G1 would produce weights more
slowly than it consumes audio, so its delay would grow over
a continuous stream. Overall, the pipeline achieves 97.65
STOI, 94.84 ESTOI, 3.676 PESQ-WB, and 20.26~dB Si-SNR at an average
power of 45.9~mW.

\subsection{System-Level Evaluation}

We consider a hearable powered by a 100~mAh, 3.7~V Li-Po battery, with an
energy budget of approximately 1330~J. As a worst-case scenario, we consider the pipeline
running continuously, with G1 working continuously. This is
pessimistic, since the delay analysis above shows that the weights can
be held over an entire utterance with no loss. Including the six
MEMS microphones (0.9~mW each, 5.4~mW total), the system draws
51.3~mW, yielding a battery
lifetime of about 7.2~h, short of the $\sim$16~h expected for
full-day use. Holding the MVDR weights and re-estimating them only once per second, while G2 continues on every frame, reduces the G1 contribution
to 8.95~mW, for a total of 44.4~mW and about 8.3~h. We notice that in this configuration, the energy budget is now dominated by the G2 executions (Table ~\ref{tab:timing_energy}).

To close the remaining gap toward full-day operation, we duty-cycle
the pipeline, activating it only during speech, which occupies no more
than approximately 30\% of the day on
average~\cite{smeds_selecting_2020}. A lightweight speech-activity
detector (SAD), executed once per G1 execution ($\sim$564~ms),
gates the remainder of the pipeline. The SAD runs in \texttt{int8} precision and requires 2.6~M
cycles per inference (29.1~MAC/cycle); quantization does not degrade
its detection performance, with the \texttt{int8} model retaining
98.5\% accuracy, 97.8\% precision, and 99.9\% recall. Microphones and
SAD remains on at all times, adding a constant power draw of 6.64~mW (5.4~mW for the microphones and 1.24~mW for the SAD). Gating is applied identically under both G1 inference schedules.
With continuous re-estimation during speech, the average draw is
20.3~mW, giving about 18.2~h of lifetime; holding the weights and re-estimating once per second lowers power consumption to 18.2~mW, about 20.3~h of lifetime. Both
configurations exceed the $\sim$16~h all-day target.




\section{Comparison with State-of-the-Art}

\begin{table*}[t]
\centering
\small
\begin{tabular}{l c l c l l c c c c}
\toprule
Model & Mpar & Quantization & QAT & Device & Deployment & ms/inf & MAC/cyc & GMAC/J & STOI\\
\midrule
TinyLSTM~\cite{TinyLSTMs}  & 0.33 & INT8 & yes & STM32F746VE & N/A & 4.26 & 0.36 & 0.14 &  N/A \\
TinyLSTM~\cite{TinyLSTMs}  & 0.46 & INT8 & yes & STM32F746VE & N/A & 2.39 & 0.36 & 0.14 &  N/A \\
\midrule
RNNoise~\cite{RNNNoise}    & 0.21 & INT8 & yes & STM32L476 & \makecell[l]{NNoM w/\\ CMSIS-NN} & 3.28 & 0.45 & 1.84 &  N/A \\
\midrule
LSTM256~\cite{Rusci}       & 1.24 & MixFP16-INT8 & no & 9-core RISC-V & GAPFlow & 2.50 & 2.11 & 17.78 &  N/A \\
GRU256~\cite{Rusci}        & 0.98 & MixFP16-INT8 & no & 9-core RISC-V & GAPFlow & 1.70 & 2.41 & 17.46 &  N/A \\
\midrule
TCN ts=4~\cite{streamease_isvlsi25} & 0.83 & INT8-BFP16 & no & 9-core RISC-V & GAPFlow & 4.3 & 3.3 & 28.90 & 93.36\\
\midrule
CNN Only~\cite{previous}   & 0.20 & MixFP32-INT8 & no & 9-core RISC-V & GAPFlow & 3.8 & 3.10 & 20.20 & 94.19\\
\textbf{Ours}              & 0.25 & MixFP32-FP16-INT8 & no & 9-core RISC-V & GAPFlow & 5 & 3.14 & 23.94 & 97.65\\
\bottomrule
\end{tabular}
\vspace{0.1cm}
\caption{Comparison with other MCU-deployed SE pipelines. }
\label{tab::references}
\end{table*}

Table~\ref{tab::references} compares our solution with state-of-the-art
speech enhancement approaches on MCUs. The baselines include
\textit{TinyLSTM}~\cite{TinyLSTMs}, evaluated on an STM32F7 MCU, and
\textit{RNNoise}~\cite{RNNNoise}, deployed on a low-power STM32L4 using
the NNoM framework with a CMSIS-NN backend~\cite{CMSISNN}. Both run on
single-core devices with 8-bit quantization, which proves effective
thanks to quantization-aware training (QAT) and model constraints that
limit the numerical range of intermediate activations. We further
compare against the GAP9-based \textit{GRU256} and \textit{LSTM256} of
Rusci et al.~\cite{Rusci}, and against the streaming TCN of Mirsalari et
al.~\cite{streamease_isvlsi25}.

Unlike the single-channel baselines, our solution processes six input
streams while retaining part of its computation in \texttt{float32}.
Despite this more demanding setting, it is more efficient than all of
them but the streaming TCN of Mirsalari et
al.~\cite{streamease_isvlsi25}: it achieves up to 7.0$\times$,
8.7$\times$, 1.30$\times$, and 1.49$\times$ higher throughput in
MAC/cycle than \textit{RNNoise}, \textit{TinyLSTM}, \textit{GRU256},
and \textit{LSTM256}, respectively, and larger gains still in energy
efficiency (GMAC/J), reaching 13.0$\times$, over two orders of
magnitude, 1.37$\times$, and 1.35$\times$ over the same baselines,
enabled by our quantization strategy together with the hardware
accelerator. The multi-timestep streaming TCN is the sole exception, exceeding our
pipeline both in utilization (3.3 vs.\ 3.14 MAC/cycle) and in energy
efficiency ($\sim$28.9 vs.\ 23.94 GMAC/J, derived from the figures
reported in~\cite{streamease_isvlsi25}). This efficiency advantage, however, comes with a single-channel formulation that cannot exploit spatial information, limiting the enhancement it can deliver. To
quantify this limitation, we reimplemented and retrained their network, the complete training pipeline not being publicly released, and evaluated it on our dataset,
obtaining 93.36 STOI, 87.87 ESTOI, 3.032 PESQ-WB, and 18.83~dB
SI-SNR, upon which our six-channel pipeline improves by +4.29 STOI,
+6.97 ESTOI, +0.64 PESQ-WB, and +1.43~dB SI-SNR. The two designs thus
occupy complementary points on the quality-efficiency trade-off.
Comparing instead with our preliminary results~\cite{previous}, which correspond to the CNN branch of our pipeline, see Section~\ref{sec:baseline},  our pipeline matches its
MAC/cycle throughput (1.01$\times$) while improving energy efficiency
by 1.19$\times$, despite the added Transformer, and achieves an
increase of +2.56~dB SI-SNR, +4.49 ESTOI, +3.46 STOI, and +0.74 PESQ-WB.

Finally, we compare our design principle against SlowFast~\cite{Slow_Fast}, which addresses the same task under an ultra-low-latency constraint but reports no MCU deployment. For this reason we restrict the comparison to the architecture.
Both systems decompose processing into a slow branch and a fast branch. The  difference lies in the timescale assigned to the slow branch.
SlowFast frames the slow branch with a window of at most 20~ms and updates it every 10~ms in its largest-reuse configuration, whereas our CNN integrates over roughly 1~s. 
The longer input sequence makes our slow branch tolerant to delay, because the CNN estimates a near-stationary representation of the acoustic scene, and its output remains valid over long intervals.
SlowFast reports monotonic degradation in enhancement quality as the reuse factor grows, because its short slow-branch window captures structure that varies on a fast timescale. 
Our system shows the opposite behavior, and Figure~\ref{fig:delay} reports the effect of the recomputation interval on STOI and SI-SNR.
Tolerance to stale beamforming weights is the property our deployment exploits, enabling a low update rate for the most expensive block in the pipeline.

\section{Conclusions}
By combining a compact CNN for mask estimation with an MVDR beamformer and a lightweight Transformer that applies a per-frame correction to the beamforming weights, and by introducing an interleaved execution strategy together with three-stage mixed-precision quantization and a lightweight SAD, we developed a multi-microphone neural beamforming pipeline that executes in real time at 24.2~mJ per inference. The hybrid architecture separates the slowly varying acoustic scene, which the CNN and MVDR part of our pipeline captures, from fast per-frame adaptation, which the Transformer handles; this separation improves enhancement quality over the standalone Transformer and the CNN + MVDR baselines. We show that the pipeline can leave the CNN-derived weights unrefreshed for long intervals, allowing the most demanding block to operate at a lower-energy point with no loss of quality. Under a realistic duty cycle ($\sim$70\% non-speech), the pipeline sustains 20~hours of continuous operation on a 100~mAh battery, including microphone power, demonstrating its practicality for all-day use in hearable devices. To the best of our knowledge, this is the first real-time, multi-channel, Transformer-based neural beamforming pipeline to run directly on an MCU.

\section*{Acknowledgment}
The authors disclose the usage of AI algorithms for grammar checking. 
\bibliography{sample}

\end{document}